\documentclass[12pt,a4paper]{article}

\usepackage{graphicx}        

\usepackage{amsmath}
\usepackage{amssymb}

\title{Nonlinear evolution wave equation for
	an artery with an aneurysm: an exact solution obtained by
	the modified method of simplest equation }
\author{E. V. Nikolova$^1$, I. P. Jordanov$^2$, 
Z. I. Dimitrova$^3$, N. K. Vitanov$^1$}

\date{$^1$Institute of Mechanics, Bulgarian Academy of Sciences, Sofia, 
Bulgaria, \\ 
$^2$ University of National and World Economy, Sofia, Bulgaria, \\
$^3$Institute of Solid State Physics, Bulgarian Academy of Sciences,
		Sofia, Bulgaria}

\begin{document}
\maketitle

\begin{abstract}
We study propagation of traveling waves in a blood filled elastic artery with an axially symmetric dilatation (an idealized aneurysm) in long-wave approximation.The processes in the injured artery are modelled by equations for the motion of the
wall of the artery and by equation for the motion of the fluid (the blood).
For the case when balance of nonlinearity, dispersion and dissipation in such a medium holds the model equations are reduced
to a version of the Korteweg-deVries-Burgers equation with variable coefficients. Exact travelling-wave
solution of this equation is obtained by the modified method of simplest
equation where the differential equation of Riccati is used as a simplest equation.
Effects of the dilatation geometry on the travelling-wave profile are considered.
\end{abstract}
\section{Introduction}
Theoretical and experimental investigation of pulse wave propagation in human arteries has 
a long history. Recently the research on blood flow has intensified because the fast
development of methods for studying nonlinear systems \cite{1}-\cite{13} allowed
to the researchers to achieve great progress  in understanding nonlinear waves in fluid-solid structures, 
such as blood-filled human arteries.
Over the past decade, however, the scientific efforts  have been concentrated on theoretical investigations of nonlinear wave propagation in arteries with a variable radius through the blood. Clearing how local imperfections appeared in an artery can disturb the blood flow can help in  predicting  the nature and main features of  various cardiovascular diseases, such as stenoses and aneurysms. In order to examine propagation of nonlinear waves in a stenosed artery, Tay and co-authors treated the artery as a homogeneous, isotropic and thin-walled elastic tube with an axially symmetric stenosis. The blood was modeled as  an incompressible inviscid fluid \cite{Tay2006}, Newtonian fluid with constant viscosity \cite{Tay2007},  and Newtonian fluid with variable viscosity \cite{Tay2008}. Using a specific perturbation method, in a long-wave approximation the authors obtained the forced Korteweg-de Vries (KdV) equation with variable coefficients \cite{Tay2006}, forced perturbed KdV equation with variable coefficients \cite{Tay2007}, and forced Korteweg-de Vries--Burgers (KdVB) equation with variable coefficients as evolution equations \cite{Tay2008}. The same theoretical frame was used in \cite{Demiray2008}, \cite{Dimitrova2015} to examine nonlinear wave propagation in an artery with a variable radius. Considering the artery as a long inhomogeneous prestretched thin elastic tube with an imperfection (presented at large by an unspecified function $f(z)$), and the blood as an incompressible inviscid fluid the authors reached again the forced KdV equation with variable coefficients. Apart from solitary propagation waves in such a system, in \cite{Dimitrova2015}, possibility of periodic waves was discussed at appropriate initial conditions. In the present work we shall focus on consideration of the blood flow through an artery with a local dilatation (an aneurysm). The aneurysm is a localized, blood-filled balloon-like bulge in the wall of a blood vessel \cite{Wikipedia}. In many cases, its rupture causes massive bleeding with associated high mortality. Motivated by investigations in \cite{Tay2006}--\cite{Dimitrova2015}, the main goal of this paper is to investigate effects of the aneurismal geometry and the blood characteristics on the propagation of nonlinear waves through an injured artery. For that purpose, we use a reductive perturbation method to obtain the nonlinear evolution equation. Exact solution of this equation is obtained by using the modified method of simplest equation. Recently, this method has been widely used to obtain general and particular solutions of economic, biological and physical models, represented by partial differential equations. The paper is organized as follows. A brief description about the derivation of equations governing the blood flow trough a dilated artery is presented in Sec. 2. In Sec. 3 we derive a basic evolution equation in long-wave approximation. A travelling wave solution of this equation is obtained in Sec. 4. Numerical simulations of the solution are presented in Sec. 5. The main conclusions based on the obtained results are summarized in Sec. 6 of the paper.

\section{Mathematical Formulation of the Basic Model}
In order to derive the model of a blood-filled artery with an aneurysm we have to consider two types equations which represent (i) the motion of the arterial wall and (ii) the motion of the blood. To model such a medium we shall treat the artery as a thin-walled incompressible prestretched hyperelastic tube with a localized axially symmetric dilatation. We shall assume the blood to be an incompressible viscous fluid. A brief formulation of the above--mentioned equations follows in the next two subsections.
\subsection{Equation of the wall}
It is well-known, that for a healthy human, the systolic pressure is about 120 mm Hg and the diastolic pressure is 80 mm Hg. Thus, the arteries are initially subjected to a mean pressure, which is about 100 mm Hg. Moreover, the elastic arteries are initially  prestretched in an axial direction. This feature minimizes its axial deformations during the pressure cycle. For example, experimental studies declare that the longitudinal motion of arteries is very small \cite{Patel}, and it is due mainly to strong vascular tethering and partly to the predominantly circumferential orientation of the elastin and collagen fibers. Taking into account these observations, and following the methodology applied in \cite{Tay2006}--\cite{Demiray2008}, we consider the artery as a circularly cylindrical tube with radius $R_{0}$, assuming that such a tube is subjected to an initial axial stretch $\lambda_{z}$ and a uniform inner pressure $P_{0}^*(Z)$. Under the action of such a variable pressure the position vector of a generic point on the tube can be described by
\begin{equation}\label{eq1}
\mathbf{r_0}=[r_{0}+f^*(z^*)]\mathbf{e_{r}}+z^*\mathbf{e_{z}},\   z^*=\lambda_{z}Z^*
\end{equation}\\
where $\mathbf{e_{r}}$ and $\mathbf{e_{z}}$ are the unit basic vectors in the cylindrical polar coordinates, $r_{0}$ is the deformed radius at the origin of the
coordinate system, $Z^*$ is axial coordinate before the deformation, $z^*$ is the axial coordinate after static deformation and $f^*(z^*)$
is a function describing the dilatation geometry. We shall specify the concrete form of $f^*(z^*)$ later. Upon the initial static deformation, we shall superimpose only a dynamical radial displacement $u^*(z^*,t^*)$, neglecting the contribution of axial displacement because of the experimental observations, given above. Then, the position vector
$r$ of a generic point on the tube can be expressed by
\begin{equation}\label{eq2}
\mathbf{r}=[r_{0}+f^*(z^*)+u^*]\mathbf{e_{r}}+z^*\mathbf{e_{z}}
\end{equation}
The arc-lengths along meridional and circumferential curves respectively, are:
\begin{equation}\label{eq3}
ds_{z}=[1+(f^{*\prime}+\frac{\partial{u^*}}{\partial{z^*}})^2]^{1/2}dz^*,\  ds_{\theta}=[r_{0}+f^*+u^*]d{\theta}
\end{equation}
In this way, the stretch ratios in the longitudinal and circumferential directions in final configuration are presented by
\begin{equation}\label{eq4}
\lambda_{1}=\lambda_{z}\Lambda, \ \lambda_{2}=\frac{1}{R_{0}}(r_{0}+f^*+u^*)
\end{equation}
where
\begin{equation}\label{eq5}
\Lambda=[1+(f^{*\prime}+\frac{\partial{u^*}}{\partial{z^*}})^2]^{1/2}
\end{equation}
The notation '$\prime$` denotes the differentiation of $f^*$ with respect to $z^*$. Then, the unit tangent vector $t$ along
the deformed meridional curve and the unit exterior normal vector $n$ to the deformed tube are expressed by
\begin{equation}\label{eq6}
\mathbf{t}=\frac{(f^{*\prime}+\frac{\partial{u^*}}{\partial{z^*}})\mathbf{e_{r}}+\mathbf{e_{z}}}{\Lambda},\quad \mathbf{n}=\frac{\mathbf{e_{r}}-(f^{*\prime}+\frac{\partial{u^*}}{\partial{z^*}})\mathbf{e_{z}}}{\Lambda}
\end{equation}
According to the assumption made about material incompressibility the following restriction holds:
\begin{equation}\label{eq7}
h=\frac{H}{\lambda_{1}\lambda_{2}}
\end{equation}\\
where $H$ and $h$ are the wall thicknesses before and after deformation, respectively. In addition, for hyperelastic materials, the tensions 
in longitudinal and circumferential directions take the form: 
\begin{equation}\label{eq8}
T_{1}=\frac{\mu^* H}{\lambda_{2}}\frac{\partial{\Pi}}{\partial{\lambda_{1}}},\ T_{2}=\frac{\mu^* H}{\lambda_{1}}\frac{\partial{\Pi}}{\partial{\lambda_{2}}}
\end{equation}\\
where $\mu^*\Pi$ is the strain energy density function of wall material as $\mu^*$ is the material shear modulus. Although the elastic properties of an injured wall section differ from those of the healthy part, here, we assume that the wall is homogeneous, i.e. $\mu^*$ is a constant through the axis $z$. Finally, according to the second Newton's law, the equation of radial motion of a small tube element placed between the planes $z^*=const$, $z^*+dz^*=const$, $\theta=const$ and
$\theta+d\theta=const$ obtains the form:
\begin{equation}\label{eq9}
-\frac{\mu^*}{\lambda_{z}}\frac{\partial{\Pi}}{\partial{\lambda_{2}}}+\mu^* R_{0}\frac{\partial}{\partial{z^*}}\left\{\frac{(f^{*\prime}+\partial{u^*}/\partial{z^*})}{\Lambda}\frac{\partial{\Pi}}{\partial{\lambda_{1}}}\right\}+\frac{P_{r}^*}{H}(r_{0}+f^*+u^*)\Lambda=\rho_{0}\frac{R_{0}}{\lambda_{z}}\frac{\partial^2{u^*}}{\partial{t^{*2}}}
\end{equation}\\
where $P_{r}^*$ is the fluid reaction force, which shall be specified later, and $\rho_{0}$ is the mass density of the tube material.
\subsection{Equation of the fluid}
Experimental studies over many years demonstrated that blood behaves as an incompressible non-Newtonian fluid because it consists of a suspension of cell formed elements in a liquid well-known as blood plasma. However, in the larger arteries (with a vessel radius  larger than 1 mm) it is plausible to assume that the blood has an approximately constant viscosity, because the vessel diameters are essentially larger than the individual cell diameters. Thus, in such vessels the non-Newtonian behavior becomes insignificant and the blood can be considered as a Newtonian fluid. 	Here, for our convenience we assume a `hydraulic approximation' and apply an averaging procedure with respect to the cross-sectional area to the Navier--Stokes equations. Then, we obtain
\begin{equation}\label{eq10}
\frac{\partial{A^*}}{\partial{t^*}}+\frac{\partial}{\partial{z^*}}(A^* \omega^*)=0
\end{equation}
\begin{equation}\label{eq11}
\frac{\partial{\omega^*}}{\partial{t^*}}+\omega^*\frac{\partial{\omega^*}}{\partial{z^*}}+\frac{1}{\rho_{f}}\frac{\partial{P^*}}{\partial{z^*}}=\frac{\mu_{f}}{\rho_{f}}\frac{\partial^2{\omega^*}}{\partial{z^{*2}}}+\frac{2 \mu_{f}}{r_{f}^2 \rho_{f}}(r\frac{\partial{V_{z}^*}}{\partial{r}})\mid_{r=r_{f}}
\end{equation}\\
where $A^*$ denotes the inner cross-sectional area, i.e., $A^* = \pi r_{f}^2$ as $r_{f}=r_{0}^*+f^*+u^*$ is the final radius of the tube after deformation. Substituting $A^*$ into Eq (\ref{eq10}) leads to
\begin{equation}\label{eq12}
2\frac{\partial{u^*}}{\partial{t^*}}+2 \omega^*[f^{*\prime}+\frac{\partial{u^*}}{\partial{z^*}}]+[r_{0}+f^*(z^*)+u^*]\frac{\partial{\omega^*}}{\partial{z^*}}=0
\end{equation} 
	We introduce the following non-dimensional quantities
\begin{eqnarray}
t^*=(\frac{R_{0}}{c_{0}})t,\ \ z^*=R_{0}z,\ \ u^*=R_{0}u,\ \ f^*=R_{0}f,\ \ \omega^*=c_{0}\omega,\ \ \mu_{f}=c_{0}R_{0}\rho_{f}\nu,\nonumber\\
P^*=\rho_{f}c_{0}^{2}p,\ \ r_{0}=R_{0}\lambda_{\theta},\ \ c_{0}^2=\frac{\mu^* H}{\rho_{f} R_{0}},\ \ m=\frac{\rho_{0} H}{\rho_{f}R_{0}}, V_{z}^*=c_{0}V_{z},\ \ r=R_{0}x\nonumber
\end{eqnarray}
and replace them into Eqs (\ref{eq12}), (\ref{eq11}) and (\ref{eq9}), respectively. Then, the final  model takes the form:
\begin{equation}\label{eq13}
2\frac{\partial{u}}{\partial{t}}+2 \omega[f^{\prime}+\frac{\partial{u}}{\partial{z}}]+[\lambda_{\theta}+f(z)+u]\frac{\partial{\omega}}{\partial{z}}=0
\end{equation}
\begin{equation}\label{eq14}			
\frac{\partial{\omega}}{\partial{t}}+\omega\frac{\partial{\omega}}{\partial{z}}+\frac{\partial{p}}{\partial{z}}=\nu\frac{\partial^2{\omega}}{\partial{z^{2}}}+\frac{2 \nu}{(\lambda_{\theta}+f+u)^2} (\frac{\partial{V_{z}}}{\partial{x}})\mid_{x=\lambda_{\theta}+f+u}
\end{equation}
\begin{eqnarray}\label{eq15}			
p=\frac{m}{\lambda_{z}(\lambda_{\theta}+f(z)+u)}\frac{\partial^2{u}}{\partial{t^2}}+\frac{1}{\lambda_{z}(\lambda_{\theta}+f(z)+u)}\frac{\partial{\Pi}}{\partial{\lambda_{2}}}\nonumber\\-\frac{1}{(\lambda_{\theta}+f(z)+u)}\frac{\partial}{\partial{z}}(\frac{f^\prime+\partial{u}/\partial{z}}{\Lambda})\frac{\partial{\Pi}}{\partial{\lambda_{1}}}+\nu\frac{(f^{\prime}+\partial{u}/\partial{z})\omega}{\lambda_{\theta}+f+u}
\end{eqnarray}
where $\lambda_{\theta}$ is the initial stretch ratio in a circumferential direction.
\section{Derivation of the evolution equation in a long--wave approximation}
In this section we shall use the long--wave approximation to employ the reductive perturbation method \cite{Jeffrey1981} for studying the propagation of small-but-finite amplitude waves in a fluid-solid structure system, presented by Eqs (\ref{eq13})--(\ref{eq15}). In the long-wave limit, it is assumed that the variation of radius along the axial coordinate is small compared with the wave length. Because this condition is valid for large arteries, next we shall treat the problem as boundary value problem. We linearize Eqs (\ref{eq13})--(\ref{eq15}), and search for their harmonic kind solution. Next, according to the obtained dispersion equation it will be appropriate to introduce the following type of stretched coordinates
\begin{equation}\label{eq16}				
\xi=\epsilon^{1/2}(z-ct),\ \ \tau=\epsilon^{3/2}z
\end{equation}
where $\epsilon$ is a small parameter measuring the weakness of nonlinearity, dispersion and dissipation, and and $c$ is a scale parameter
which will be determined from the solution. Then, $z=\epsilon^{-3/2} \tau$, and $f(\epsilon^{-3/2} \tau)=\epsilon \chi(\xi,\tau)$. Thus, the variables $u, \omega$ and $p$ are functions of the variables $(\xi,\tau)$ and the small parameter $\epsilon$. Taking into account the effect of dilatation, we assume $f$ to be of order of 5/2, i.e.
\begin{equation}\label{eq17}
\chi(\xi,\tau)=\epsilon h(\tau)
\end{equation}
In addition, taking into account the effect of viscosity, the order of viscosity is assumed to be $O(1/2)$, i.e. 
\begin{equation}\label{eq18}
\nu=\epsilon^{1/2}\overline{\nu}
\end{equation}
The last assumption ensures balance of nonlinearity, dispersion and dissipation in the system. In order to eliminate the unknown term $(\frac{\partial{V_{z}}}{\partial{x}})$ in Eq (\ref{eq14}), we use the transformation \cite{Bakirtas2003}
\begin{equation}\label{eq19}
y=(\lambda_{\theta}+\epsilon h(\tau)+u-x)\epsilon^2
\end{equation}
For the long wave limit, it is assumed that the field quantities may be expanded into asymptotic series as
\begin{eqnarray}\label{eq20}			
u=\epsilon u_{1}+\epsilon^2 u_{2}+...,\ \omega=\epsilon \omega_{1}+\epsilon^2 \omega_{2}+..,\nonumber\\
p=p_{0}+\epsilon p_{1}+\epsilon^2 p_{2}+...,\ \lambda_{1}\cong \lambda_{z},\nonumber\\
\lambda_{2}=\lambda_{\theta}+\epsilon (u_{1}+h)+\epsilon^2 (u_{2}+(u_{1}+h)^2)+...,\ \frac{1}{\lambda_{\theta}\lambda_{z}}\frac{\partial{\Pi}}{\partial{\lambda_{z}}}=\gamma_{0}\\
\frac{1}{\lambda_{\theta}\lambda_{z}}\frac{\partial{\Pi}}{\partial{\lambda_{\theta}}}=\beta_{0}+\beta_{1}(u_{1}+h)\epsilon+(\beta_{1}u_{2}+\beta_{2}(u_{1}+h)^2)\epsilon^2+..\nonumber
\end{eqnarray}
where
\begin{eqnarray}\label{eq21}
\beta_{0}=\frac{1}{\lambda_{\theta}\lambda_{z}}\frac{\partial{\Pi}}{\partial{\lambda_{\theta}}},\ \ \beta_{1}=\frac{1}{\lambda_{\theta}\lambda_{z}}\frac{\partial^2{\Pi}}{\partial{\lambda_{\theta}^2}},\ \ \beta_{2}=\frac{1}{2 \lambda_{\theta}\lambda_{z}}\frac{\partial^3{\Pi}}{\partial{\lambda_{\theta}^3}}
\end{eqnarray}
Substituting (\ref{eq16})--(\ref{eq20}) into Eqs (\ref{eq13})--(\ref{eq15}), we obtain the following differential sets:\\

\textit{O ($\epsilon$) equations}

\begin{equation}\label{eq22}				
-2c\frac{\partial{u_{1}}}{\partial{\xi}}+\lambda_{\theta}\frac{\partial{\omega_{1}}}{\partial{\xi}}=0,\  -c\frac{\partial{\omega_{1}}}{\partial{\xi}}+\frac{\partial{p_{1}}}{\partial{\xi}}=0,\   p_{1}=\gamma_{1}(u_{1}+h)
\end{equation}

\textit{O ($\epsilon^2$) equations}

\begin{eqnarray}\label{eq23}				
-2c \frac{\partial{u_{2}}}{\partial{\xi}}+2\omega_{1} \frac{\partial{u_{1}}}{\partial{\xi}}+\lambda_{\theta}\frac{\partial{\omega_{2}}}{\partial{\xi}}+[u_{1}+h]\frac{\partial{\omega_{1}}}{\partial{\xi}}+\lambda_{\theta}\frac{\partial{\omega_{1}}}{\partial{\tau}}=0\nonumber\\
-c\frac{\partial{\omega_{2}}}{\partial{\xi}}+\omega_{1}\frac{\partial{\omega_{1}}}{\partial{\xi}}+\frac{\partial{p_{2}}}{\partial{\xi}}+\frac{\partial{p_{1}}}{\partial{\tau}}-\overline{\nu} \frac{\partial^2{\omega_{1}}}{\partial{\xi}^2}=0\\
p_{2}=(\frac{m c^2}{\lambda_{\theta}\lambda{z}}-\gamma_{0})\frac{\partial^2{u_{1}}}{\partial{\xi}^2}+\gamma_{1} u_{2}+\gamma_{2}(u_{1}+h)^2\nonumber\
\end{eqnarray}
From Eqs (\ref{eq22}), we obtain
\begin{equation}	\label{eq24}			
u_{1}=U(\xi, \tau),\ \ \omega_{1}=\frac{2c}{\lambda_{\theta}}U, \ \ p_{1}=\frac{2c^2}{\lambda_{\theta}}U+\gamma_{1} h
\end{equation}
as $\gamma_1=\frac{2c^2}{\lambda_{\theta}}$. We introduce (\ref{eq24}) into (\ref{eq23}), and obtain
\begin{equation}\label{eq25}				
-2 c \frac{\partial{u_{2}}}{\partial{\xi}}+\frac{4c}{\lambda_{\theta}}U\frac{\partial{U}}{\partial{\xi}}+\lambda_{\theta}\frac{\partial{\omega_{2}}}{\partial{\xi}}+2c \frac{\partial{U}}{\partial{\tau}}+\frac{2c}{\lambda_{\theta}}(U+h)\frac{\partial{U}}{\partial{\xi}}=0
\end{equation}
\begin{equation}\label{eq26}				
-c \frac{\partial{\omega_{2}}}{\partial{\xi}}+\frac{4c^2}{\lambda_{\theta}^2}U\frac{\partial{U}}{\partial{\xi}}+\frac{2c^2}{\lambda_{\theta}}\frac{\partial{U}}{\partial{\tau}} +\gamma_{1} h^{\prime}+\frac{\partial{p_{2}}}{\partial{\xi}}-\frac{4c^2}{\lambda_{\theta}^2}\frac{\partial^2{U}}{\partial{\xi}^2}=0
\end{equation}
\begin{equation}\label{eq27}		
p_{2}=(\frac{m c^2}{\lambda_{\theta}\lambda{z}}-\gamma_{0})\frac{\partial^2{U}}{\partial{\xi}^2}+\gamma_{1} u_{2}+\gamma_{2}U^2+\gamma_{2}h(\tau)U+\gamma_{2}h(\tau)^2
\end{equation}
Replacing Eq (\ref{eq27}) into Eq (\ref{eq26}), and eliminating $\omega_2$ between Eq (\ref{eq25}) and Eq (\ref{eq26}), the final evolution equation takes the form:
\begin{equation}\label{eq28}
\frac{\partial{U}}{\partial{\tau}}+\mu_{1}U\frac{\partial{U}}{\partial{\xi}}-\mu_{2}\frac{\partial^2{U}}{\partial{\xi}^2}+\mu_{3}\frac{\partial^3{U}}{\partial{\xi}^3}+\mu_{4}(\tau)\frac{\partial{U}}{\partial{\xi}}+\mu(\tau)=0
\end{equation}
where
\begin{eqnarray}\label{eq29}		
\mu_{1}=\frac{5}{2\lambda_{\theta}}+\frac{\gamma_{2}}{\gamma_{1}},\ \ \mu_{2}=\frac{\overline{\nu}}{\lambda_{\theta}},\ \ \mu_{3}=\frac{m}{4 \lambda_{z}}-\frac{\gamma_{0}}{2 \gamma_{1}},\\
\mu_{4}(\tau)=h(\tau)(\frac{1}{2 \lambda_{\theta}}+\frac{\gamma_{2}}{\gamma_{1}}),\ \ \mu(\tau)=\frac{1}{2} h^{\prime}(\tau)\nonumber\
\end{eqnarray}
and
\begin{equation}\label{eq30}		
\gamma_{1}=\beta_{1}-\frac{\beta_{0}}{\lambda_{\theta}},\ \gamma_{2}=\beta_{2}-\frac{\beta_{1}}{\lambda_{\theta}}
\end{equation}
Finally we have to objectify the idealized aneurysm shape. For an idealized abdominal aortic aneurysm (AAA), $h(\tau)=\delta exp(\frac{-\tau^2}{2 L^2})$, where $\delta$ is the aneurysm height, i.e. $\delta=r_{max}-r_{0}$, and $l$ is the aneurysm length \cite{Gopalakrishnan2014}. In order to normalize these geometric quantities, we non-dimensionalize  $\delta$ by the inlet radius (diameter). Then, the non-dimensional coefficient can be presented by $\delta^{\prime}=DI-1$, where $DI=2r_{max}/2r_{0}=D_{max}/D_{0}$ is a geometric measure of AAA, which is known as a diameter index or a dilatation index \cite{Raut2013}. In the same manner, the aneurysm length $L$ is normalized by the maximum aneurysm diameter ($D_{max}$), i.e. $l^{\prime}=L/D_{max}=1/SI$, where $SI$ is a ratio, which is known as a sacular index of AAA \cite{Raut2013}. For AAAs, $D_{max}$ varies from 3 cm to 8.5 cm, and $L$ varies from 5 cm to 10--12 cm.

\section{Analytical solution for the nonlinear evolution equation: Application of the modified method of simplest equation}
In this section we shall derive a progressive wave solution for the variable coefficients evolution equation, presented by (\ref{eq28}).
We shall make change of the function and the variables in the the evolution equation with variable coefficients as follows:\\
Let us introduce $U(\xi,\tau)=V(\xi,\tau)+\phi(\tau)$. Replacing this substitution into (\ref{eq28}), leads to:
\begin{equation}\nonumber
\frac{\partial{V}}{\partial{\tau}}+\mu_{1}(V-\int{\mu(\tau)}d \tau) \frac{\partial{V}}{\partial{\xi}}-\mu_{2}\frac{\partial^2{V}}{\partial{\xi}^2}+\mu_{3}\frac{\partial^3{U}}{\partial{\xi}^3}+\mu_{4}(\tau)\frac{\partial{V}}{\partial{\xi}}=0.
\end{equation}
Now, we introduce the coordinate transformation
\begin{equation}\nonumber
\tau^{\prime}=\tau,\ \xi^{\prime}=\xi-\int[\mu_{4}(\tau)+\mu_{1}\int{\mu(\tau)}d \tau]d \tau
\end{equation}
Then, Eq (\ref{eq28}) is reduced to the generalized KdVB equation:
\begin{equation}\label{eq31}
\frac{\partial{V}}{\partial{\tau}^{\prime}}+\mu_{1}\frac{\partial{V}}{\partial{\xi}^{\prime}}-\mu_{2}\frac{\partial^2{V}}{\partial{\xi}^{\prime 2}}+\mu_{3}\frac{\partial^3{U}}{\partial{\xi}^{\prime 3}}=0.
\end{equation}
Next, we shall find an analytical solution of Eq (\ref{eq31}) applying the modified method of simplest equation \cite{v1}--\cite{v10}. This method has its roots in the research of Kudryashov and ther authors \cite{y1}-\cite{y5}. The short description of the modified method of simplest equation is as follows. First of all by means of an appropriate ansatz (for an example the traveling-wave ansatz) the solved  of nonlinear partial differential equation for the unknown function $\eta$ is reduced to a nonlinear ordinary differential equation that includes $\eta$ and its derivatives with respect to the
traveling wave coordinate $\zeta$
\begin{equation}\label{eq32}
\Phi \left( \eta, \eta_{\zeta},\eta_{\zeta \zeta},\dots \right) = 0
\end{equation}
Then the finite-series solution 
\begin{equation}\label{eq33}
\eta(\zeta) = \sum_{\mu=-\kappa}^{\kappa_1} a_{\mu} [g (\zeta)]^{\mu}
\end{equation}
is substituted in (\ref{eq32}). $a_\mu$ are coefficients and $g(\zeta)$ is solution of
simpler ordinary differential equation called simplest equation. Let the result
of this substitution be a polynomial of $g(\zeta)$. Eq. (\ref{eq33}) is a solution of Eq.(\ref{eq32}) 
if all coefficients of the obtained polynomial of $g(\zeta)$ are equal to $0$. This condition 
leads to a system of nonlinear algebraic equations. Each nontrivial solution of the last  system  leads to a solution of the studied  
nonlinear partial differential equation. In addition, in order to obtain the solution of Eq.(\ref{eq32}) by the above method we have
to ensure that each coefficient of the obtained polynomial of $g(\zeta)$ contains at least two terms. To
do this within the scope of the modified method
of the simplest equation we have to balance the
highest powers of $g(\zeta)$ that are obtained from the
different terms of the solved equation of kind
(\ref{eq32}). As a result of this we obtain an additional
equation between some of the parameters of the
equation and the solution. This equation is called a
balance equation.
\par
Introducing the transformation $\zeta = \xi^{\prime} -v\tau^{\prime}$, we search for solution of (\ref{eq31})  of kind $V=V(\zeta)={\sum\limits_{r=0}^q}  a_{r}g_{\zeta}^{r}$, where $g^{\prime}_{\zeta}={\sum\limits_{j=0}^{m}} b_{j}g^{j}$. Here $a_{r}$ and $b_{j}$ are parameters, and $g(\zeta)$ is a solution of some ordinary differential equation, referred to as the simplest equation. The balance equation is $q=2m-2$. We assume that $m=2$, i.e. the  equation of Riccati will play the role of simplest equation. Then
\begin{equation}\label{34}
V=a_{0}+a_{1}g+a_{2}g^2,\ \ \frac{dg}{d\zeta}=b_{0}+b_{1}g+b_{2}g^2
\end{equation}
The differential equation of Riccati can be written as
\begin{equation}\label{eq35}
\left(\frac{dg}{d \zeta} \right)^2 = c_0 +c_1 g+c_2 g^2+c_3 g^3+c_4 g^4
\end{equation}
where
\begin{equation}\label{eq36}
c_0=b_0^2; \ c_1=2b_0 b_1; \ c_2=2b_0 b_2 + b_1^2; \ c_3=2b_1 b_2; \ c_4=b_2^2
\end{equation}
and its solutions are given in \cite{v3}. The relationships among the coefficients of the solution and the coefficients of the model are derived by solving a system of five algebraic equations, and they are
\begin{eqnarray}\label{eq37}
a_{0}=-\frac{1}{25}\frac{-3 \mu_{2}^2-30 \mu_2 \mu_3 b_1+75 \mu_3^2 b_1^2+25 v \mu_3}{\mu_1 \mu_3};\nonumber\\
a_1=-\frac{12}{5}\frac{b_2 (5 \mu_3 b_1-\mu_2)}{\mu_1};\quad a_2=-12 \frac{\mu_3 b_2^2}{\mu_1};\quad b_0=\frac{1}{100} \frac{25 \mu_3^2 b_1^2-\mu_2^2}{b_2 \mu_3^2}
\end{eqnarray}
Here $b_{1},b_{2}$ are free parameters. Then, the solution of the evolution equation with constant coefficients (Eq (\ref{eq31})) is
\begin{eqnarray}\label{eq38}
V(\zeta)=-\frac{1}{25}\frac{-3 \mu_{2}^2-30 \mu_2 \mu_3 b_1+75 \mu_3^2 b_1^2+25 v \mu_3}{\mu_1 \mu_3}-\\
-\frac{12}{5}\frac{b_2 (5 \mu_3 b_1-\mu_2)}{\mu_1}g(\zeta)-12 \frac{\mu_3 b_2^2}{\mu_1} g(\zeta)^2\nonumber
\end{eqnarray}\\
where
\begin{eqnarray}\label{eq39}
g(\zeta)=-\frac{b_{1}}{2b_{2}}-\frac{\Delta}{2b_{2}} \tanh{(\frac{\Delta (\zeta+\zeta_{0})}{2})}+\\
+\frac{\exp{(\frac{\Delta (\zeta+\zeta_{0})}{2})}}{2 \cosh(\frac{\Delta (\zeta+\zeta_{0})}{2})\frac{b_2}{\Delta}+2 C^*\exp(\frac{\Delta (\zeta+\zeta_{0})}{2})\cosh(\frac{\Delta (\zeta+\zeta_{0})}{2})}\nonumber
\end{eqnarray}
In Eq. (\ref{eq39}) $\Delta=\sqrt{b_{1}^2-4b_{0}b_{2}}>0$, and $\xi_0$ and $C^*$ are constants of integration.
The solution of the evolution equation with variable coefficients (Eq (\ref{eq28})) is
\begin{equation}\label{eq40}
U(\xi,\tau)=V(\zeta)-\int{\mu(\tau)}d\tau
\end{equation}
where
\begin{equation}\label{eq41}
\zeta=\xi-v\tau-\int[\mu_{1}\int{\mu(\tau)}d \tau+\mu_{4}(\tau)]d \tau
\end{equation}
\section{Numerical findings and discussions}
\begin{figure}
			\centering
			\includegraphics[width=0.6\textwidth]{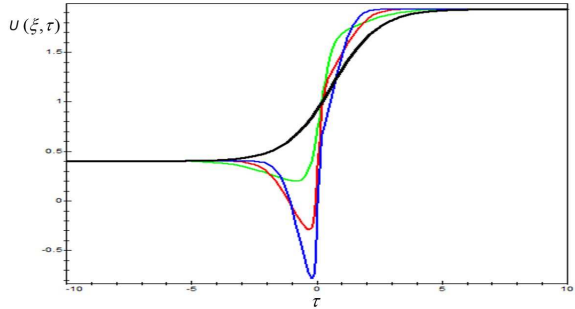}
			\caption{Wave profiles of the radial displacement $U$ at $D_{max}$=2 cm (the black line); $D_{max}$=3 cm (the green line); $D_{max}$=5 cm (the red line); $D_{max}$=7 cm (the blue line) ($L=5 cm$)}
			\label{U_1}
		\end{figure}

In order to see the effect of dilatation on the wave profiles of investigated quantities, we need the values of coefficients $\beta_0, \beta_1,\beta_2,\gamma_0,\gamma_1,\gamma_2,\mu_1,\mu_2,\mu_{4}(\tau)$ and $\mu(\tau)$. For that purpose, the constitutive relation for tube material must be specified. Here, unlike \cite{Tay2006}--\cite{Dimitrova2015}, we assume that the arterial wall is an incompressible, anisotropic and hyperelastic material. The mechanical behaviour of such a material can be defined by the strain energy function of Fung for arteries \cite{Fung1993}:
\begin{equation}\label{eq42}
\Pi=C(e^{Q}-1),\quad  Q=C_{1}E_{QQ}^2+C_{2}E_{ZZ}^2+2 C_{3}E_{QQ}E_{ZZ}
\end{equation}
where $E_{QQ}$ and $E_{ZZ}$ are the Green--Lagrange strains in the circumferential and axial directions, respectively, and $C,C_1,C_2,C_3$ are material constants. Taking into account that $E_{QQ}=1/2(\lambda_{\theta}^2-1)$ and  $E_{ZZ}=1/2(\lambda_{z}^2-1)$, we substitute (\ref{eq42}) in (\ref{eq21}) and (\ref{eq30}), and obtain:
\begin{eqnarray}\label{eq43}
&&\beta_{0}=\frac{1}{\lambda_{z}}(\frac{C_1}{2}+C_{3}(\lambda_{z}^2-1))F(\lambda_{\theta}\lambda_{z})\nonumber\\ 
&&\beta_{1}=\frac{1}{\lambda_{z}\lambda_{\theta}}(\frac{C_1}{2}+C_{3}(\lambda_{z}^2-1))(1+\lambda_{\theta}^2(\frac{C_1}{2}+C_{3}(\lambda_{z}^2-1)))F(\lambda_{\theta}\lambda_{z})\nonumber\\
&&\beta_{2}=\frac{1}{2 \lambda_{z}}(\frac{C_1}{2}+C_{3}(\lambda_{z}^2-1))^2(3+\lambda_{\theta}^2(\frac{C_1}{2}+C_{3}(\lambda_{z}^2-1)))F(\lambda_{\theta}\lambda_{z})\\
&&\gamma_{0}=\frac{1}{\lambda_{\theta}}(\frac{C_2}{2}+C_{3}(\lambda_{\theta}^2-1))F(\lambda_{\theta}\lambda_{z}),\ 
\gamma_{1}=\frac{1}{\lambda_{z}}(\frac{C_1}{2}+C_{3}(\lambda_{z}^2-1))^2F(\lambda_{\theta}\lambda_{z}),\nonumber\\
&&\gamma_{2}=\frac{1}{\lambda_{z}}(\frac{C_1}{2}+C_{3}(\lambda_{z}^2-1))(\frac{\lambda_{\theta}^2}{2}(\frac{C_1}{2}+
C_{3}(\lambda_{z}^2-1))^2+\frac{5}{2}(\frac{C_1}{2}+C_{3}(\lambda_{z}^2-1))\nonumber\\
&&-\frac{1}{\lambda_{\theta}^2})F(\lambda_{\theta}\lambda_{z})\nonumber
\end{eqnarray}
where
\begin{equation}\label{eq44}
F(\lambda_{\theta}\lambda_{z})=C \exp(\frac{C_{1}}{4}(\lambda_{\theta}^2-1)+\frac{C_{2}}{4}(\lambda_{z}^2-1)+\frac{C_3}{2}(\lambda_{\theta}^2-1)(\lambda_{z}^2-1))
\end{equation}\\
The numerical values of material coefficients in (\ref{eq44}) are as follows: $C=2.5,C_1=14.5,C_2=7,C_3=0.1$. They were derived in \cite{Avril2010} from experimental data of  human aortic wall segments applying a specific inverse technique. Assuming the initial deformation $\lambda_{z}=1.5,\lambda_{\theta}=1.2$, we obtain the following values for the coefficients: $\beta_0= 554.97,\beta_1=5374,\beta_2= 
27872.89,\gamma_0=333.36, \gamma_1= 4911.52,\gamma_2= 23394.55
$. Then, the numerical values of the coefficients in Eq. (\ref{eq28}) are:
\begin{eqnarray}
\mu_1= 6.85;\ \mu_2= 0.42;\ \mu3= -0.017;\\
 \mu_4(\tau)=5.36 \delta^{\prime} \exp(-\tau^2/2 l^{\prime 2}), ;\ \mu(\tau)=-\delta^{\prime}\tau \exp(-\tau^2/2 l^{\prime 2})/2 l^{\prime 2}.\nonumber
		\end{eqnarray}
Using these numerical values, the travelling-wave solution of Eq.(\ref{eq28}) 
for various time $\xi$ ($\xi$=0..5 (years)) is plotted in Fig (\ref{U_1}). It is seen that in absence of arterial dilatation (at $D_{max}=D_0$) the wave demonstrates a pure kink (the black line in the figure). In the case of an aneurismal artery, however, a slight wave drop, followed by a prompt wave jump is observed (see the other lines in the figure). In addition, the wave drop region diminishes when the value of the maximal aneurysm diameter ($D_{max}$) decreases. This result seems to be admissible from the point of view of arterial mechanics.
		
\section{Conclusions}
Modelling the injured artery as a thin-walled prestetched, anisotropic and hyperelastic tube with a local imperfection (an aneurysm), and the blood as a Newtonian fluid we have derived an evolution equation for propagation of nonlinear waves in this complex medium. Numerical values of the model parameters are determined for concrete mechanical characteristics of the arterial wall and concrete aneurismal geometry. We have obtained an analytical solution of the evolution equation of a travelling-wave type. The numerical simulations of this solution demonstrate that for a healthy artery the wave is of a pure kink type, while wave drop and jump effects are observed when a local arterial dilatation arises. 

\section{Acknowledgement}
	This work was supported by the UNWE project for 
	scientific researchers with grant agreement No. NID NI-- 21/2016 and the Bulgarian National Science Fund with grant agreement No. DFNI I 02-3/
	12.12.2014.		


\begin{thebibliography}{99.}
\bibitem{1} 
Hilborn~RC. (2000) Chaos and Nonlinear Dynamics. Oxford University Press, New York
\bibitem{2}
Strogatz~SH. (2015) Nonlinear Dynamics and Chaos. Westview Press, Boulder, CO.
\bibitem{3}
Dimitrova~ZI, Vitanov~NK. (2000) Influence of adaptation on the nonlinear dynamics of a system of competing populations. Physics Letters A 272, 368 - 380.
\bibitem{4} Dimitrova~ZI, Vitanov~NK. (2001) Dynamical consequences of adaptation
of the growth rates in a system of three competing populations. Journal of Physics A: Mathematical and General 34, 7459 - 7473.
\bibitem{5} 
Dimitrova~ZI, Vitanov~NK. (2001) Adaptation and its impact on the dynamics of a system of three competing populations. Physica A 300, 91 - 115.
\bibitem{6}
Boeck T, Vitanov NK. (2002) Low-dimensional chaos in zero-Prandtl-number Benard-Marangoni 
convection. Physical  Review E 65 (3), 037203.
\bibitem{7}
Dimitrova~ZI, Vitanov~NK. (2004) Chaotic pairwise competition. Theoretical Population 
Biology 66, 1 - 12.
\bibitem{8}
Vitanov~NK, Sakai~K, Jordanov~IP, Managi~S, Demura~K. (2007)
Analysis of a Japan government intervention on the domestic agriculture
market. Physica A 382, 330 - 335.
\bibitem{9}
Panchev~S, Spassova~T, Vitanov~NK. (2007)
Analytical and numerical investigation of two families of Lorenz-like dynamical systems. Chaos, Solitons \& Fractals 33, 1658 - 1671.
\bibitem{10}
Vitanov~NK, Sakai~K, Dimitrova~ZI. (2008) SSA, PCA, TDPSC, ACFA:
Useful combination of methods for analysis of short and nonstationary time series.
Chaos, Solitons \& Fractals 37, 187 - 202.
\bibitem{11}
Vitanov~NK, Dimitrova~ZI, Ausloos~M. (2010) Verhulst-Lotka-Volterra (VLV) model of ideological struggle. Physica A 389, 4970 - 4980.
\bibitem{12}
Vitanov~NK, Ausloos~M, Rotundo~G. (2012)
Discrete model of ideological struggle accounting for migration. Advances in Complex 
Systems 15 (supp01) 1250049.
\bibitem{x1} 
Nikolova~E, Jordanov~I, Vitanov~NK. (2013) Dynamical features of the 
quasi-stationary microRNA-mediated protein translation process supported by 
eIF4F translation initiation factors. Computers \& Mathematics with Applications 66,
1716 - 1725.
\bibitem{x3}
Vitanov~NK, Vitanov~KN. (2014) Population dynamics in presence of state dependent fluctuations. Computers \& Mathematics with Applications 68, 962 - 971.
\bibitem{x4}
Vitanov~NK, Hoffmann~NP, Wernitz~B. (2014) Nonlinear time series analysis of vibration data from a friction brake: SSA, PCA, and MFDFA. Chaos, Solitons \& Fractals
69, 90 - 99.
\bibitem{x2}
Nikolova~E, Goranova~E, Dimitrova~Z. (2016) Assessment of rupture risk factors of abdominal aortic aneurysms in Bulgarian patients using a finite element based system. Comptes 
rendus de l'Académie bulgare des Sciences 69, 1213 - 1222.
\bibitem{13}
Vitanov~NK. (2016)
Science Dynamics and Research Production: Indicators, Indexes, Statistical Laws and Mathematical Models. Springer International.
	\bibitem{Tay2006}
	Tay~KG. (2006) Forced Korteweg-de Vries equation in an elastic tube filled with an inviscid fluid. International Journal of Engineering Science 44, 621--632.
	\bibitem{Tay2007}
	Tay~KG, Ong~CT and Mohamad~MN (2007) Forced perturbed Korteweg-de Vries equation in an elastic tube filled with a viscous fluid. International Journal of Engineering Science 45, 339--349.
	\bibitem{Tay2008}
	Tay~KG, Demiray~H. (2008) Forced Korteweg-de Vries–Burgers equation in an elastic tube filled with a variable viscosity fluid. Chaos, Solitons \& Fractals 38, 1134–-1145.
	\bibitem{Demiray2008}
	Demiray H. (2008) Non-linear waves in a fluid-filled inhomogeneous elastic tube with variable radius. International Journal of Non-Linear Mechanics 43, 241--245.
	\bibitem{Dimitrova2015}
	Dimitrova~ZI. (2015) Numerical investigation of nonlinear waves connected to blood flow in an elastic tube of variable radius. Journal of Theoretical and Applied Mechanics 45, 79--92.
	\bibitem{Wikipedia} Aneurysm, From Wikipedia, the free encyclopedia, https://en.wikipedia.org/wiki/Aneurysm
	\bibitem{Patel}
	Patel~PJ, Greenfield~JC, Fry~DL. (1964) In vivo pressure length radius relationship in certain blood vessels in man and dog, in: E.O. Attinger
	(Ed.), Pulsatile Blood Flow, McGraw-Hill, New York, p. 277.
	\bibitem{Jeffrey1981}
	Jeffrey~A, Kawahara~T. (1981) Asymptotic methods in nonlinear wave theory. Pitman, Boston.
	\bibitem{Bakirtas2003}
	Bakirtas~I, Antar~N (2003) Evolution equations for nonlinear waves in a tapered elastic tube filled with a viscous fluid. International Journal of Engineering Science 41, 1163–1176.
	\bibitem{v1}
	Vitanov~NK, Dimitrova~ZI, Kantz~H. (2010) Modified method of simplest equation and its application to nonlinear PDEs.  Applied Mathematics and Computation 216, 
	2587 - 2595.
	\bibitem{v2}
	Vitanov~NK. (2010) Applications of simplest equations of Bernoulli and Riccati kind for obtaining exact traveling-wave solutions for a class of PDEs with polynomial nonlinearity. Communications in Nonlinear Science and Numerical Simulation 15, 
	2050 - 2060.
	\bibitem{v3}
	Vitanov~NK, Dimitrova~ZI. (2010) Application of the method of simplest equation for obtaining exact traveling-wave solutions for two classes of model PDEs from ecology and population dynamics. Communications in Nonlinear Science and Numerical 
	Simulation 15, 2836 - 2845.
	\bibitem{v4}
	Vitanov~NK. (2011) Modified method of simplest equation: Powerful tool for obtaining exact and approximate traveling-wave solutions of nonlinear PDEs. Communications in Nonlinear Science and Numerical Simulations \textbf{16}, 1176 - 1185.
	\bibitem{v5}
	Vitanov~NK. (2011) On modified method of simplest equation for obtaining exact and approximate solutions of nonliear PDEs: the role of simplest equation. Communications in Nonlinear Science and Numerical Simulations 16, 4215 - 4231.
	\bibitem{v6}
	Vitanov~NK, Dimitrova~ZI, Vitanov~KN. (2011) On the class of nonlinear 
	PDEs that can be treated by the modified method of simplest equation. 
	Application to the generalized Degasperis-Processi equation and b-equation. Communications in Nonlinear Science and Numerical Simulation 16, 1176 - 1185.
	\bibitem{v7}
	Vitanov~NK, Dimitrova~ZI, Vitanov~KN. (2013) Traveling waves and statistical distributions connected to systems of interacting populations. Computers \& Mathematics with Applications 66, 1666 - 1684.
	\bibitem{v8}
	Vitanov~NK, Dimitrova~ZI, Kantz~H. (2013) Application of the method of 
	simplest equation for obtaining exact traveling-wave solutions for the 
	extended Korteweg-deVries equation and generalized Camassa-Holm equation.
	Applied Mathematics and Computation 219, 7480 - 7492.
	\bibitem{v9}
	Vitanov~NK, Dimitrova~ZI. (2014) Solitary wave solutions for nonlinear 
	partial differential equations that contain monomials of odd and even 
	grades with respect to partiacipating derivatives Applied Mathematics and 
	Computation 247, 213 - 217.
	\bibitem{v10}
	Vitanov~NK, Dimitrova~ZI, Vitanov~KN. (2015) Modified method of simplest equation for obtaining exact analytical solutions of nonlinear partial differential equations: further development of the methodology with applications.  Applied Mathematics and Computation 269, 363 - 378.
	\bibitem{y1}
	Kudryashov~NA. (2005) Simplest equation method to look for exact solutions of nonlinear differential equations. Chaos, Solitons \& Fractals 24, 1217 - 1231.
	\bibitem{y2}
	Kudryashov~NA, Loguinova~NB. (2008) Extended simplest equation method for nonlinear differential equations. Applied Mathematics and Computations 205, 396 - 402.
	\bibitem{y3}
	Martinov~N, Vitanov~N. (1992) On some solutions of the two-dimensional sine-Gordon equation. Journal of Physics A: Mathematical and General 25, L419 - L426. 
	\bibitem{y4}
	Martinov~N, Vitanov~N. (1992) New class of running-wave solutions of the (2+1)-dimensional sine-Gordon equation. Journal of Physics A: Mathematical and General 27,
	4611 - 4618.
	\bibitem{yy}
	Martinov~N, Vitanov~N. (1992) Running wave solutions of the two-dimensional sine-Gordon equation 25, 3609 - 3613. 
	\bibitem{y5}
	Vitanov~NK.(1996) On travelling waves and double-periodic structures in two-dimensional sine-Gordon systems. Journal of Physics A: Mathematical and General 29,
	5195 - 5207.
	\bibitem{Fung1993}
	Fung~ Y. (1993) Biomechanics: Mechanical Properties of Living Tissues. New York, Springer.
	\bibitem{Avril2010}
	Avril~S, Badel~P, Duprey~ A. (2010) Anisotropic and hyperelastic identification of in vitro human arteries from full-field optical measurements. J Biomech  43, 2978-2985.
	\bibitem{Gopalakrishnan2014}
	Gopalakrishnan~SS, Benoît~P, Biesheuvel~A. (2014) Dynamics of pulsatile flow through model abdominal aortic aneurysms. J. Fluid Mech. 758, 150--179.
	\bibitem{Raut2013}
	Raut~SS, Chandra~S, Shum~J, Finol~EA. (2013) The role of geometric and biomechanical factors in abdominal aortic aneurysm rupture risk assessment. Ann Biomed Eng. 41, 1459-1477.
	
\end{thebibliography}
\end{document}